\documentclass[letterpaper, 10 pt, conference]{ieeeconf}  
\usepackage{xcolor} 
\usepackage{url}

\IEEEoverridecommandlockouts                              

\usepackage{graphicx} 
\usepackage{tcolorbox}
\usepackage{CJKutf8} 
\usepackage{xcolor} 

\title{\LARGE \bf
Exploring EEG Indicators to Evaluate Listening Difficulties \\in Noisy Environments}

\author{Azuki Onaya$^{1}$ and Hiroki Tanaka$^{1}$
\thanks{*This work was funded by the Japan Society for the Promotion of Science (grant 22K12151).}
\thanks{$^{1}$Azuki Onaya and Hiroki Tanaka are with College of Liberal Arts Division of Arts and Sciences,
        International Christian University, Mitaka, Japan.
      }}%

\begin{document}

\maketitle
\thispagestyle{empty}
\pagestyle{empty}

\begin{abstract}
Auditory processing difficulties involve challenges in understanding speech in noisy environments despite normal hearing. However, the neural mechanisms remain unclear, and standardized diagnostic criteria are lacking. This study examined neural indicators using EEG under realistic noisy conditions. Ten Japanese-speaking university students participated in auditory tasks, including a resting state, a lecture attention task with background noise, and a task requiring attention to background noise. The study analyzed the peak frequency and power of alpha waves, the long-range temporal correlation of alpha oscillations, and the absolute power of delta waves. Results showed a significant reduction in the power of alpha waves during the background noise attention task, suggesting increased cognitive load. In contrast, the peak frequency of alpha waves remained stable, indicating limited sensitivity to cognitive demand changes. Long-range temporal correlation increased under tasks requiring auditory attention, reflecting sustained attention-related neural dynamics, while the absolute power of delta waves showed no significant variation across tasks. Regression analysis revealed a significant negative correlation between the power of alpha waves in noisy conditions and screening scores for auditory processing difficulties, suggesting its potential as a neural indicator. Although no statistically significant relationship was found between screening scores and learning style preferences, a weak trend suggested a possible preference for visual learning among individuals with higher screening scores for auditory processing difficulties. These findings improve the understanding of auditory processing difficulties' neural basis and highlight the potential role of electroencephalography-based indicators in their evaluation. Future studies with larger sample sizes and advanced analysis methods are needed to strengthen these findings' validity and clinical application.

\end{abstract}

\section{Introduction}

Listening difficulties or auditory processing disorder (APD) are conditions characterized by difficulties in understanding speech in noisy environments \cite{ASHA_CAPD}. Difficulties in listening were observed among the general public.
Despite its clinical significance, the underlying mechanisms of APD remain insufficiently understood, and no standardized diagnostic criteria or universally accepted treatment protocols have been established \cite{Alanazi2023}. The symptoms of APD exhibit substantial inter-individual variability, encompassing deficits in auditory discrimination, selective attention, and phonological awareness  \cite{stewart2017, Barry2015}. These deficits can have profound implications for academic achievement, professional performance, and social communication \cite{ASHA2005, Zoppo2015}.

Prior studies have suggested that individuals with APD may exhibit a predisposition toward visual dominance in information processing \cite{obuchi2016}. This observation raises the possibility of a relationship between APD and learning style preferences, particularly in the context of the VARK model, which categorizes learners based on their preferred sensory modalities: visual, auditory, reading/writing, and kinesthetic, abbreviated as V, A, R, and K, respectively \cite{vark_introduction}. 
Furthermore, electroencephalography (EEG) has been widely utilized to investigate neural correlates of auditory processing and cognitive functions \cite{fraxa2025, stavrinos2018}. Several EEG-based indices, including alpha peak frequency (APF) and alpha peak power (APP), have been linked to cognitive load and attentional regulation \cite{eqlimi2019, angelakis2004, zhang2021, deng2019}. Additionally, long-range temporal correlation (LRTC) and delta absolute power (DAP) have been proposed as EEG indicators associated with sustained attention and auditory processing efficiency \cite{linkenkaer2001, palva2013, stam2007, eqlimi2019}.

The present study aims to elucidate the neurophysiological mechanisms underlying APD by examining EEG indicators in relation to APD screening scores and learning style preferences. Specifically, we investigate neural activity patterns associated with auditory attention in noisy environments and explore whether individuals exhibiting higher APD screening scores demonstrate a compensatory reliance on visual learning strategies. 
Under these objectives, the study formulates the following hypotheses. 

\begin{enumerate}
  \item Neural characteristics related to auditory processing are expected to vary across different auditory tasks in noisy environments. Specifically, APF is predicted to decrease during lecture attention tasks in multi-talker noise, reflecting increased cognitive demands. Similarly, reductions in APP and DAP are anticipated, indicating greater neural effort for auditory processing and impaired auditory tracking capabilities. Additionally, LRTC within the alpha band is expected to decline in conditions requiring sustained auditory attention.
  \item It is hypothesized that individuals with higher APD screening scores will show a preference for visual learning, as assessed by the VARK questionnaire. This preference is expected to be associated with distinct neural activity patterns, including reduced APP, APF, and DAP, indicating lower auditory processing efficiency. These findings aim to enhance the understanding of the neural mechanisms underlying APD and its connection to individual differences in learning style preferences.
\end{enumerate}

\section{Methods}

This section outlines the methods, including participant information, experimental procedures, and statistical analysis.

\subsection{Participants}
Ten Japanese-speaking university students (aged 20–28) without hearing disorders participated in this study. Informed consent was obtained from all participants prior to the experiment.

\subsection{Experimental Protocol}
Participants performed three consecutive auditory tasks while keeping their eyes closed to minimize eye-related muscle movement artifacts. The first task, the Resting Task, involved no auditory stimuli and served as a baseline condition. The Lecture Attention Task required participants to focus on lecture content embedded in background noise, where the lecture and noise were presented at equal sound pressure levels via Creative Pebble X SP-PBLX-BKA PC speakers. After completing the Lecture Attention Task, participants took a comprehension test consisting of 10 questions. The Background Attention Task involved attending to salient auditory events (e.g., phone ringtones) embedded within the background noise, using the same speaker system to ensure consistent audio quality across tasks.

\subsection{Stimuli}
For the Lecture Attention Task, auditory stimuli comprised artificially generated lecture content created using ChatGPT-4o to eliminate any advantage from prior knowledge. The generation prompt instructed the creation of a five-minute fictional lecture along with a corresponding comprehension test tailored for experimental conditions involving background noise.

The lecture materials were pre-recorded using an AKG Ara-Y3 USB condenser microphone, featuring a 24-bit/96kHz sampling rate and a wide frequency response range (20Hz–20kHz). The lecture and background noise were mixed in Audacity, ensuring that both components maintained equal sound pressure levels. To simulate real-world noisy environments, salient auditory events, such as phone ringtones, were randomly embedded within the background noise.
The original lecture content and comprehension test were initially written in Japanese and later translated into English for presentation in this paper, as illustrated in Fig.~\ref{tab:elysium_text} and Table~\ref{tab:elysium_quiz}.

\begin{figure}[htbp]
\vspace{5pt}
    \centering
    \begin{tcolorbox}[
        colframe=black, 
        colback=yellow!10!white, 
        coltitle=black, 
        sharp corners, 
        width=0.95\linewidth,  
        boxrule=0.8pt,
        arc=2mm,
        left=5pt, right=5pt, top=5pt, bottom=5pt,
        before skip=5pt, 
    ]
        ``Elysium" is a mysterious substance that affects the mind. When inhaled in small amounts, it enables people to instantly share memories and emotions with one another. The emergence of this revolutionary substance had the potential to fundamentally change the nature of communication. Scientists and thinkers were highly optimistic about the new means of interaction it could provide. It was believed that by using Elysium, individuals could transcend their emotions and experiences, forming a ``network of empathy" where everyone was interconnected and lived in mutual understanding. In particular, Elysium was expected to serve as a key to fostering a peaceful society by eliminating the emotional disconnect that often leads to misunderstandings and conflicts. Captivated by this groundbreaking technology, the protagonist, Kim Ji-hoon, dedicated his life to researching Elysium. He believed that the division of human consciousness was the root cause of conflict, misunderstanding, and opposition. As long as individuals maintained independent awareness, true understanding and empathy would remain unattainable, making the realization of a peaceful society difficult. However, Kim was convinced that if all people could share a single collective consciousness through Elysium, humanity would be able to unify emotions and memories, eradicating misunderstandings and resentment. His project, ``New Consciousness," aimed to distribute Elysium widely, allowing society as a whole to share thoughts and emotions, ultimately achieving genuine understanding and lasting peace...
    \end{tcolorbox}
    \caption{Fictional Lecture Material: ``Elysium"}
    \label{tab:elysium_text}
\end{figure}

\begin{table}[!t]
    \caption{Multiple-choice questions from the comprehension test in the Lecture Attention Task}
    \label{tab:elysium_quiz}
    \begin{enumerate}
    \item \mbox{What effect does Elysium have?}
    
    \begin{tabular}{p{4.8cm} p{2.2cm}}
        \hline
        \textbf{Options} & \textbf{Correct/Incorrect} \\
        \hline
        Controls people's thoughts & Incorrect \\
        Allows sharing of memories and emotions & \textbf{Correct} \\
        Enhances physical abilities & Incorrect \\
        Enables prediction of the future & Incorrect \\
        \hline
    \end{tabular}
    \\
    \\
    \item \mbox{ What is the protagonist's name?}
    \begin{tabular}{p{4.8cm} p{2.2cm}}
        \hline
        \textbf{Options} & \textbf{Correct/Incorrect} \\
        \hline
        Lee Jun & Incorrect \\
        Jeon Min-ho & Incorrect \\
        Park Seung-yeon & Incorrect \\
        Kim Ji-hoon & \textbf{Correct} \\
        \hline
    \end{tabular}
    \\
    \\
    \item \mbox{What is the name of the project planned by the protagonist?}
    \begin{tabular}{p{4.8cm} p{2.2cm}}
        \hline
        \textbf{Options} & \textbf{Correct/Incorrect} \\
        \hline
        Elysium Plan & Incorrect \\
        Mind Linking & Incorrect \\
        New Consciousness & \textbf{Correct} \\
        Mental Think Tank & Incorrect \\
        \hline
    \end{tabular}
    \end{enumerate}
\end{table}

In the Background Attention Task, participants were exposed to background noise featuring salient auditory events. To maintain consistency, the same multi-talker babble used in the Lecture Attention Task was utilized. All auditory stimuli were delivered through speakers, ensuring consistent sound quality across all conditions.

\subsection{Data Collection}
EEG signals were continuously recorded using the CGX Quick-32r system from 32 scalp locations, including the standard 10-20 system and ten additional positions. The EEG recording was conducted in an electromagnetically shielded soundproof room to minimize external noise and interference. Participants listened to monaural audio playback presented via the speaker positioned one meter away. EEG data were referenced to Cz and sampled at 2048 Hz.

Event timing during the experiment was manually recorded to align with task conditions and auditory stimuli. For the Resting Task, start and end times were manually marked using a timer. In the Lecture Attention and Background Attention Tasks, the onset and offset of auditory stimuli were recorded manually using a numeric keypad. This approach ensured precise segmentation of EEG data corresponding to the task periods, enabling accurate preprocessing and analysis.

\subsection{Questionnaires}

\subsubsection{Auditory Processing Disorder Screening Score}
APD tendencies were assessed using the questionnaire developed by Obuchi and Kaga \cite{obuchi2019}, which integrates elements from the Speech, Spatial, and Qualities of Hearing Scale-12 (SSQ-12) \cite{SSQ12} and the Hearing Questionnaire 2002 \cite{hearing2002}. The SSQ-12 evaluates auditory processing across three domains:
\begin{itemize}
    \item \textbf{Speech hearing}: Assessing the ability to understand speech in complex auditory environments.
    \item \textbf{Spatial hearing}: Assessing spatial auditory abilities, including sound localization and movement perception.
    \item \textbf{Hearing quality}: Assessing aspects such as sound segregation, clarity, and listening effort.
\end{itemize}

Additionally, four items from the Hearing Questionnaire 2002 were incorporated to assess behavioral and emotional responses to listening difficulties, such as:
\begin{itemize}
    \item \textbf{Behavioral reactions (2 items)}: Assessing responses to listening challenges in daily life.
    \item \textbf{Emotional reactions (2 items)}: Assessing emotional impact, such as frustration or stress.
\end{itemize}

This hybrid questionnaire offers a multidimensional evaluation of auditory processing difficulties, making it well-suited for assessing APD tendencies within the context of the present study.

\subsubsection{Learning Style}
The VARK (Visual, Auditory, Reading/Writing, Kinesthetic) questionnaire was administered to assess participants' learning style preferences \cite{vark_questionnaire}. This questionnaire includes multiple-choice questions aimed at assessing the sensory modalities participants primarily rely on for information processing. Participants were asked to select all applicable options for each question to capture their multimodal preferences.

The questionnaire consists of 16 situational items, such as preferences for:
\begin{itemize}
    \item Website types based on key features (e.g., detailed descriptions, interactive elements, or visual appeal).
    \item Methods for giving directions (e.g., drawing maps, providing written instructions, or accompanying someone).
    \item Receiving feedback after an exam (e.g., verbal feedback, written results, or visual graphs).
    \item Approaches to learning new skills (e.g., studying manuals, analyzing diagrams, or engaging in hands-on practice).
\end{itemize}

Responses were scored based on the frequency of preferences for each modality (V, A, R, or K). Participants were categorized as single-mode or multimodal learners according to their responses, offering insights into individual differences in learning styles and their connection to auditory processing tendencies.

The VARK questionnaire was selected for its clarity and practical applicability across research contexts. Unlike other learning style inventories that assess a wide range of attributes, VARK specifically focuses on sensory modalities, aligning with the study’s aim of exploring the relationship between auditory processing tendencies and learning preferences.

\subsection{Data Analysis}
\subsubsection{EEG Pre-processing}
EEG data preprocessing was conducted using the EEGLAB toolbox in MATLAB \cite{Delorme2004}. First, the raw EEG data were downsampled from 2048 Hz to 512 Hz to reduce computational load while maintaining temporal resolution. A bandpass filter (0.5–134 Hz) was applied using a finite impulse response filter with a Hamming windowed sinc function. The EEG signals were re-referenced to the central electrode Cz to reduce reference bias and improve signal quality. The average across all electrodes was then computed for subsequent analysis.

\subsubsection{Statistical Analysis}
To investigate the relationship between APD tendencies and EEG indicators, statistical tests were conducted for each measure.
For group comparisons, the Wilcoxon rank-sum test was used to assess differences in APP and LRTC across task conditions.
To examine associations between EEG features and APD screening scores, multiple linear regression analysis was performed, with EEG indicators (e.g., APP) as independent variables and APD scores as the dependent variable.

All statistical analyses were conducted using Python, with data handling performed via Pandas and NumPy, statistical tests carried out using SciPy.stats and Statsmodels, and significance thresholds adjusted as needed.

\section{Results}

First, no statistically significant relationship was found between screening scores and learning style preferences. However, a weak trend indicated a potential preference for visual learning among individuals with higher screening scores for auditory processing difficulties. Besides, this section presents the results of significant EEG indicators related to APD.


\subsection{Alpha Peak Power}
As Table~\ref{tab:app_summary} and Table~\ref{tab:app_wilcoxon} show, a statistically significant reduction in APP was observed in the background attention condition relative to the resting condition ($p = 0.0027$), indicating an increase in cognitive demand when processing salient background sounds amidst noise. The finding suggests that auditory environments with competing stimuli require greater cognitive effort, potentially reflecting the neural mechanisms involved in selective auditory attention.

\begin{table}[!t]
    \centering
    \caption{Mean and Variance for APP}
    \label{tab:app_summary}
    \begin{tabular}{lcc}
    \hline
    \textbf{Task} & \textbf{Mean} & \textbf{Variance} \\
    \hline
    Resting Task & 11.00 & 56.00  \\
    Lecture Attention Task & 10.31 & 58.29  \\
    Background Attention Task & 9.26 & 42.53  \\
    \hline
    \end{tabular}
\end{table}

\begin{table}[!t]
    \centering
    \caption{Wilcoxon Rank Sum Test for APP}
    \label{tab:app_wilcoxon}
    \begin{tabular}{lcc}
    \hline
    \textbf{Comparison} & \textbf{$p$-value} \\ 
    \hline
    Resting vs Lecture Attention & $0.19$  \\
    Resting vs Background Attention & $0.0027$  \\
    Lecture Attention vs Background Attention & $0.18$  \\ 
    \hline
    \end{tabular}
\end{table}

\subsection{Long-Range Temporal Correlation}
Significant increases in LRTC were observed across both auditory attention conditions, highlighting its sensitivity to sustained attention demands as shown in Table~\ref{tab:lrtc_summary} and Table~\ref{tab:lrtc_wilcoxon}. Notably, comparisons between all task conditions revealed significant differences, suggesting that neural coordination dynamics are modulated by auditory attention requirements.

\begin{table}[!t]
    \centering
    \caption{Mean and Variance for LRTC}
    \label{tab:lrtc_summary}
    \begin{tabular}{lccc}
    \hline
    \textbf{Task} & \textbf{Mean} & \textbf{Variance} \\
    \hline
    Resting & 0.65 & 0.021  \\
    Lecture Attention & 0.70 & 0.033 \\
    Background Attention & 0.73 & 0.025  \\
    \hline
    \end{tabular}
\end{table}

\begin{table}[!t]
    \centering
    \caption{Wilcoxon Rank Sum Test for LRTC}
    \label{tab:lrtc_wilcoxon}
    \begin{tabular}{lc}
    \hline
    \textbf{Comparison} & \textbf{$p$-value} \\
    \hline
    Resting vs Lecture Attention & $9.40 \times 10^{-3}$ \\
    Resting vs Background Attention & $6.37 \times 10^{-10}$ \\
    Lecture Attention vs Background Attention & $2.18 \times 10^{-3}$ \\
    \hline
    \end{tabular}
\end{table}

\subsection{Regression Analysis with APD Scores}
Multiple linear regression analysis revealed a significant negative association between APF and APD screening scores in the background attention condition ($p = 0.04$, $R^2 = 0.67$; see Table~\ref{tab:noise_only_regression}). This finding suggests that higher APF values correspond to lower APD scores, highlighting APF as a potential neural indicator for APD.

\begin{table}[!t]
\centering
\caption{Regression Results for the Background Attention Task}
\label{tab:noise_only_regression}
\begin{tabular}{lcccc}
\hline
\textbf{Variable} & \textbf{Estimate} & \textbf{S.E.} & \textbf{$t$-value} & \textbf{$p$-value} \\
\hline
APP & -2.26 & 1.53 & -1.48 & 0.20 \\
APF & -16.58 & 5.92 & -2.80 & \textbf{0.04} \\
LRTC & 129.92 & 68.67 & 1.89 & 0.12 \\
DAP & -0.00049 & 0.00063 & -0.78 & 0.47 \\
\hline
\end{tabular}
\end{table}

\section{Discussion}
These findings suggest that APP and APF may serve as potential EEG indicators of APD screening scores, particularly in environments with competing auditory stimuli. The observed reduction in APP under noisy conditions is consistent with prior research showing that decreased alpha power reflects increased cognitive demand \cite{eqlimi2019}. Meanwhile, the relative stability of APF across tasks indicates its limited sensitivity to environmental fluctuations, yet its predictive value for APD screening scores remains significant \cite{angelakis2004}.

Moreover, the sensitivity of LRTC to task conditions underscores its role in neural coordination during auditory attention \cite{linkenkaer2001}. The absence of significant findings for DAP and learning styles suggests that these measures may require larger sample sizes or more sensitive tasks to elucidate meaningful relationships.

In this paper, the small sample size ($n$ = 10) restricts the generalizability of the findings. Furthermore, the experimental design may not have induced sufficient cognitive load variations to reveal subtle neural differences. Additionally, the reliance on self-reported questionnaires could limit the accuracy of APD and learning style assessments.

\section{Conclusion}
This study advances our understanding of the neural correlates of APD screening scores by integrating EEG analysis with behavioral assessments. APP and APF emerge as promising indicators for evaluating auditory processing difficulties in noisy environments. Future research involving larger, more diverse samples and advanced analysis techniques is recommended to validate and extend these findings.


\subsection{Acknowledgment}
Funding for this study was provided by the Japan Society for the Promotion of Science (grant number 22K12151).

\bibliographystyle{ieeetr}
\bibliography{root}

\end{document}